\documentstyle[epsf,psfig,rotate,amstex,amssymb,rotate]{l-aa} 

\begin{document}

   \thesaurus{08         
	       08.01.3;  
               08.03.2;  
               08.05.4;  
               08.18.1;  
             }

\title{New $\lambda \ Bootis$ stars with a shell}

\author{B. Hauck\inst{1} \and D. Ballereau\inst{2} \and J. Chauville\inst{2}}

\offprints{B. Hauck}

\institute{Institut d'Astronomie de l'Universit\'e de Lausanne,
 CH-1290 Chavannes-des-Bois, Switzerland
 \and
 Observatoire de Paris-Meudon, DASGAL, F-92125 Meudon Principal Cedex, France
}
\date{Received ; accepted }

\maketitle

\markboth{B. Hauck et al.: New $\lambda \ Bootis$ stars with a shell}
{B. Hauck et al.: New $\lambda \ Bootis$ stars with a shell}

\begin{abstract}
We publish here the second part of our spectroscopic survey at high dispersion of some known and suspected $\lambda \ Bootis$ stars with a view to detecting circumstellar shell features. Eight stars of our sample exhibit such  features. These stars are fast rotators, a result which is in line with Holweger and Rentzsch-Holm's study (1995). The analysis of the photometric data has allowed us to confirm the exclusion of a few stars misclassified from the $\lambda \ Bootis$ group.
\end{abstract}

\section{Introduction}
We have undertaken a spectroscopic survey at high dispersion of 
some known $\lambda \ Bootis$ stars with a view to testing the hypothesis of 
Venn and Lambert (1990), who suggested a depleted gas accretion 
model. In a first paper we found (Hauck et al. 1995, hereafter Paper 
I) circumstellar shell (CS) features in the CaII-K line of two stars (HD 
16955 and HD 204965) and a possible such feature for HD 220061. 
New observations were made at the Haute-Provence Observatory in 
1994 and 1995 and we present here evidence for the presence of 
circumstellar components around some other $\lambda \ Bootis$ stars.

Since publication of Paper I, many papers on the $\lambda \ Bootis$ stars 
have been published, among which we may mention those from 
Holweger and Rentzsch-Holm (1995) and Andrillat et al. (1995). 
Holweger and Rentzsch-Holm report on high-resolution observations 
of the CaII-K line in two samples of A stars, $\lambda \ Bootis$ and dusty normal 
A stars, finding CS components for five out of eleven  $\lambda \ Bootis$ stars. 
Their conclusion is that the presence of detectable amounts of 
circumstellar gas in A stars is rare among chemically normal A 
stars. However the  $\lambda \ Bootis$ stars are metal-deficient and nearly half of their sample  are  surrounded by circumstellar gas. A similar conclusion is 
reached by Andrillat et al. (1995), who observed a sample of 20  
 $\lambda \ Bootis$ stars in three wavelength regions in the near infrared. Seven 
stars of this sample exhibit evidence of shells. It is also important to 
pay attention to temporal variations of $V_r$ for both the stellar and 
circumstellar lines. Many authors have reported lately on the 
pulsation and variability of some  $\lambda \ Bootis$ stars (Weiss et al. 1994, 
Kuschnig et al. 1994a, b, Paunzen et al. 1995a, b, Paunzen 1995, 
Bohlender et al. 1996) and Holweger and Rentzsch-Holm (1995) have 
mentioned a $V_r$ variation of the CS component of HD 111786 between 
1990 and 1994. At the earlier date the CS component is slightly blue 
shifted, while at the second date it is red shifted. Faraggiana et al. (1997) consider this star to be a spectroscopic binary whose $\lambda \ Bootis$ primary has a high rotation speed, while the secondary has an F type and a slow rotation; one of  the two components pulsates.

\section{Observations and reduction}
\subsection{Observations}

The first difficulty we encounter in a study on the $\lambda \  Bootis$ stars is 
that of establishing a list of candidates. Not all the $\lambda \ Bootis$ 
classifications are based on the same criteria, although Gray (1988) has 
proposed a clear working definition. The stars selected for our survey 
come from those recognised as being $\lambda \  Bootis$ stars according to 
spectroscopic criteria. Our first sources were Gray (1988) and Gray \& 
Garrison (1987, 1989a, b). We subsequently used three other papers by Abt 
(1984), Abt \& Morrell (1995) and Andrillat et al. (1995). Such a choice 
can obviously reveal some disagreement between these authors, our Table 1 
showing that many stars considered as $\lambda \ Bootis$ by Abt or Abt and Morrell are 
not classified $\lambda \  Bootis$ by Gray and Garrison.

According to Gray (1988) some stars were classified in the literature as 
$\lambda \  Bootis$ on the basis of a weak MgII $\lambda$ 4481 line, but some other groups 
of stars display the same feature, as for example many Ap and helium-weak 
stars. Thus we can understand why some stars classified $\lambda \  Bootis$ by Abt 
are classified Ap by other authors.

The above discussion shows the difficulty in selecting real $\lambda \  Bootis$ 
stars. Some authors (Baschek et al. 1984, Faraggiana et al. 1990) have 
used UV criteria, but due to the small number of stars in their sample it 
is not sure if they really are a general property of $\lambda \  Bootis$ stars. 
Faraggiana (1987) well resumed the situation before the work of Gray 
(1988), while Gerbaldi and Faraggiana (1993) have produced a list of what 
we could call real $\lambda \  Bootis$ stars. However, such a list would now need to 
be revised by adding new stars.

It may be seen from Table 1 of Gerbaldi and Faraggiana (1993) that among the 
stars of our Table 1 the following are considered to be real $\lambda \  Bootis$ 
stars:  HD 31295, HD 38545, HD 110411, HD 111604, HD 125162 and HD 
221756. A survey of the literature allows us to propose some more stars:  
HD 36496, HD 39283 from Andrillat et al. (1995), while HD 217782 was 
proposed by Parenago (1958) as a $\lambda \ Bootis$ star and this property was 
confirmed by Andrillat et al.

It will be seen in section 3 that we propose the exclusion of four stars 
from the $\lambda \  Bootis$ group:  HD 47152, HD 108283, HD 204965 and HD 225180. 
HD 159082 and HD 196821 are also classified as HgMn stars and, taking 
into account the remark of Gray (1988), they may also be excluded. The 
status of HD 112097 is not very clear. This star is classified as Am by 
Levato and Abt (1978) and is classified F0Vp ($\lambda \  Bootis$, met: A7) by Abt 
and Morrell (1995). Photometric data (see section 3) agree with $\lambda \  Bootis$ 
type, this star being slightly metal-deficient. HD 110377 is found to be 
a $\delta \  Sct$ star by Peniche et al. (1981).

The remaining stars have either an alternate classification or are 
classified only by Abt (1984) or Abt and Morrell (1995). Gray and Garrison (1987, 1989a, b) 
have classified the following stars as normal, while Abt or Abt and Morrell 
classified them as $\lambda \  Bootis$:  HD 11503, HD 16811, HD 30739, HD 118623, HD 
125489, HD 153808, HD 161868, HD 210418, HD 214454 and HD 220061.

HD 34787 is classified as $\lambda \  Bootis$ by Abt and Morrell (1995), while Faraggiana 
et al. (1990) have excluded it. Finally HD 2904, HD 5789 and HD 109980
are classified as $\lambda \  Bootis$ only by Abt and Morrell and 
HD 141851 by Abt. Such an 
enumeration of disagreement clearly shows the necessity for a critical 
survey by a spectroscopic specialist of all $\lambda \ Bootis$ stars and $\lambda \ Bootis$ 
candidates. Since the submission of this paper, a consolidated catalogue of $\lambda \ Bootis$ stars has been lately published (Paunzen et al. 1997) and surely it could play this role.

Some of these stars are already mentioned in the 
literature as having circumstellar matter. They are HD 38545 
(Stürenburg 1993, Bohlender and Walker 1994, Paper I), HD 39283 
(Andrillat et al. 1995), HD 108283 (Jaschek et al. 1991) and HD 217782 
(Andrillat et al. 1995). Concerning HD~31295, HD~110411 and HD~221756, Stürenburg (1993)
mention the presence of a marginal shell. For some stars of our sample an IR excess is 
mentioned in the literature. King (1994) has found such an excess for 
HD 31295, HD 125162 and HD 161868 and a marginal excess for HD 
111604. Cheng et al. (1995) find an IR excess for HD 110411.

Our observations were made during four sessions at the Haute-Provence Observatory in November 1994, February 1995, May 1995 
and October-November 1995, using the 1.52 m telescope equipped 
with the AURELIE spectrograph. The detector is a double-element 
TH7832 with sets of 2048 photodiodes of 750 x 13 mm. Two gratings 
were used ($N^o$ 7 with 1800 lines/mm and $N^o$ 2 with 1200 lines/mm). 
More details on AURELIE can be found in Gillet et al. 1994).

Table 1 lists the stars observed during the four sessions, together 
with some of their characteristics. The fourth column, S1, gives the source of $\lambda \  Bootis$ classification.  Spectral types are from Abt and Morrell (column 6) and Gray and Garrison (column 7). Radial velocities are from the Bright Star Catalogue 
(Hoffleit and Jaschek, 1982), while $Vsin\ i$ is taken from Abt and 
Morrell (1995) and photometric data, including $m_v$, are taken from 
the Geneva Database. Table 2 gives a record of the new observations, 
the latter being made in the following spectral regions: $H\gamma$, CaII-K, 
NaI-D and for a few stars, at $H\alpha$ and $H\beta$.

\subsection{Radial velocities of photospheric lines}
As in Paper I, we have taken advantage of an excellent signal-to-noise 
ratio to determine the radial velocity of the main photospheric lines 
($H\alpha$, $H\beta$, $H\gamma$, $H_8$ and CaII-K). It is recalled that two procedures were 
used to derive the radial velocity. Each line was measured twice: by 
parabolic fit from the bottom of the profile and by the gravity centre 
of its lower part, limited by a straight horizontal line. Table 3 gives 
these two measurements for each observed line. In this table, radial 
velocities are given to one decimal digit, for the following reason. The 
wavelength of the line's bottom is defined by the minimum of the parabola 
fitted on $N\pm 7$ point in the line core, and its precision is given 
approximately by the formula:
\begin{equation}
\Delta X \simeq \frac{2}{\sqrt{N-2}} \frac{sampling\ step}{S/N}
\end{equation}
obtained by a similar way as the formula given by Gerstenkorn et al. (1977).
With a 0.1 \AA\ step for grating $N^o 2$ and S/N $\geq$ 300, we therefore obtain $\Delta X \simeq 3 \ 10^{-4}$ \AA, which is lower than the error of the Th-Ar dispersion curve, 0.015 \AA. For a $\lambda$ of 4000 \AA (Ca II), this represents, in RV of the parabolic fit, a precision of $\sim$ 1 $km\ s^{-1}$, thus justifying the 1-decimal digit of RVs in Table 3.

In general, the radial velocities of the hydrogen lines are in 
agreement, although they are sometimes different from those of CaII-
K. In paper I, we suggested that these differences may be explained 
either by the different velocities of the hydrogen winds and of ionised 
calcium (stratification of the elements) or by the presence of a binary 
system whose spectral components vary in phase opposition, the hottest component showing H-lines and the coolest Ca II-K line. In some 
cases our mean radial velocities show a marked difference with those 
given by Hoffleit and Jaschek (1982). We will revert to these points in 
section 5.

\subsection{Radial velocities of shell lines}
All the line profiles were examined carefully to detect a CS 
component. In addition to HD 16955 and HD 204965, already 
discussed in paper I, we found eight more stars with such a 
component. Figure 1 shows the profile of the CaII-K line for these 
stars. It may be noted that no clearly pronounced CS component was 
found for HD 39283, while Andrillat et al. (1995) consider this star as 
LB shell. The radial velocity and the equivalent width of the CS 
component were then determined. Table 4 gives for each star these 
two quantities.

We obtained a spectrum of the CaII-K line of the $\lambda \ Bootis$ prototype 
star (HD 125162), which deserves special mention. Barring errors on 
our part, no CaII-K line profiles of HD 125162 have been published up to the 
present.

Figure 2 shows this line, between $H_{\epsilon}$ and $H_8$, for $\lambda \ Boo$ and HD 
31295 ($\pi^1$ Ori), which has the same spectral type (A0V) and the same 
projected rotation velocity $(Vsin\ i=105\ km\ s^{-1})$. The equivalent CaII-K 
widths are respectively:
\begin{flushleft}
$W_{CaII-K} = 400.8$ m\AA\ $(\lambda \ Boo)$
\end{flushleft}
\begin{flushleft}
$W_{CaII-K} = 637.6$ m\AA\ (HD 31295)
\end{flushleft}
Figure 3 shows an enlargement of the $\lambda \ Boo$ CaII-K line illustrated 
in Figure 2.

Contrary to what one would expect, this star has no circumstellar 
component in fine absorption, but what would appear to be two slight 
components in emission. It will be seen that at the bottom of the CaII-K line of $\lambda Boo$ two weak emissions that are well separated from each other appear above the photospheric profile.They may be compared with what appears on the $H_\gamma$ line of the Be star 
HD 47054 in Ballereau et al. 1995 (p. 440). However, this emission was 
only observed on one profile and has to be confirmed by further 
observations.

\section{Photometric data}
Data for table 1 are taken from the Geneva Database maintained by G. 
Burki. B2-V1 is a temperature parameter, d a luminosity parameter, 
 $\Delta m_2$ a blanketing parameter and $\Delta$ a parameter sensitive to luminosity 
and blanketing. Before using B2-V1 to determine $T_{\mathrm eff}$, we have to 
consider if the stars are reddened. For this we can first of all use the 
Strömgren photometry and then the code developed by Künzli et al. (1997). 
Only HD 225180 was detected as being reddened. 

As regards a luminosity effect  on B2-V1, Hauck (1985) has shown that as far as classes V to III are concerned the $T_{\mathrm eff}$ vs B2-V1 relation is not influenced by luminosity effect.
$T_{\mathrm eff}$ was derived from a relation between $T_{\mathrm eff}$ and 
B2-V1 (Hauck 1994). The precision of such a $T_{\mathrm eff}$ is of order of $\pm\  200\ K$. Hauck \& Künzli (1996) have shown that the $T_{\mathrm eff}$ determined by the relation between $T_{\mathrm eff}$ and B2-V1 have an accurency of the same order as those determined by the code of Künzli et al. (1997), i.e. $\pm\  200\ K$.
Since many $\lambda \ Bootis$ stars have a high $Vsin\ i$ value
it should be recalled that B2-V1 is affected by such high values 
(Hauck and Slettebak 1989). Stars with high $Vsin\ i$ values have too red 
a B2-V1 value. For A1 stars for instance, the mean B2-V1 value for 
$Vsin\ i$ values between 0 and 100 is 0.134 while it is -0.100 for $Vsin\ i$ 
values between 200 and 400. Thus for the latter stars the $T_{\mathrm eff}$ value 
given in Table 1 is only an estimate of the real $T_{\mathrm eff}$.

Since the $\lambda \ Bootis$ stars are  metal-deficient, it is interesting 
to use the properties of the Geneva system to consider whether the 
stars of our sample meet this assumption. $\Delta m_2$, the difference between $m_2(stars)-m_2(Hyades)$, is a blanketing parameter, well-correlated  to [Fe/H] (Hauck 1978). Stars with a negative  $\Delta m_2$ value have a lower [Fe/H] than the Hyades stars. According to $\Delta m_2$, the following stars are metal-deficient: HD 
36496, HD 110377, HD 111604, HD 141851, a few others being only 
slightly metal-deficient (-0.028 $\leq$  $\Delta m_2$ $\leq$ -0.018). 

Some other stars possess interesting photometric properties:

\noindent
HD 108283 has a positive value of  $\Delta m_2$ (of the same order as an Am 
or Ap star) and we may assume that this star does not belong to the $\lambda Bootis$ group. It is mentioned as Sr? by Renson (1992), classified as A9 
IVnp Sr II by Gray and Garrison (1989) but A9 Vp ($\lambda \ Boo$) by Abt and 
Morrell (1995). Jaschek et al. (1991) found a moderate shell for this 
star. We have also found shell components for the same star.

\noindent
HD 47152 is a well-known $Hg$ star (Osawa 1965) and this is confirmed by its 
positive value of the $\Delta$(V1-G) parameter.

\noindent
HD 204965 has high d and $\Delta$ values and thus in a d vs $\Delta$ diagram it 
is located in the region of luminosity class III stars. Iliev and Barzova 
(1995) have recently confirmed the assumption that $\lambda \ Bootis$ stars are 
main-sequence stars. So we may wonder if HD 204965 belongs to the 
$\lambda \ Bootis$ group. We can also reach the same conclusion for HD 225180, 
which is classified by Gray and Garrison (1987) as A1 II-III and this 
is confirmed by our d and $\Delta$ parameters. This star is reddened, 
therefore we cannot derive a  $\Delta m_2$ value.

Thus, on the basis of the photometric data, we can exclude these 
four stars from the $\lambda \ Bootis$ group.

\section{Stars with circumstellar component}
As mentioned in section 2.3, we have found eight stars, in addition to HD 
16955 and HD 204965 already discussed in paper I, presenting a CS 
component and we have used a relation proposed by Higurashi and Hirata 
(1978) to derive the extension of the shell. This relation is:
\begin{equation}
r/r^*=(\Delta V_{1/2}/Vsin\ i)^{1/(1+j)}
\end{equation}
where $r/r*$ is the extension of the shell in stellar radius,
$\Delta V_{1/2}$ the half-width at half-depth and
$j$ a parameter depending from the rotation law of the shell gas. 
In our case, keplerian rotation, $j = 1/2$.

We have applied it to the stars of our sample with a CS component, 
Table 4 showing the results of our calculations.

In addition to the stars of table 4, we have found two stars for 
which the presence of a CS component is less marked:  HD 39283 
and HD 108283. On the basis of its photometric data HD 108283 is not 
a $\lambda \ Bootis$ star but an Am or Ap star. No evidence has been found for a 
CS component in the case of HD 31295 and HD 110411, whereas 
Sturenburg (1993) finds a marginal CS component for both of 
these stars. For HD 110411, Holweger and Rentzsch-Holm (1995) find 
a very weak CS component.

The presence of a CS component in the stars of table 4 is 
confirmed by some other studies. Andrillat et al. (1995) also find a CS 
component for HD 38545 and HD 217782. Furthermore, 
Sturenburg (1993) and Bohlender \& Walker (1994) indicate the 
presence of a CS component for HD 38545. The fact that some 
authors find a CS component and others do not is perhaps due to 
a temporal variation of this CS component. In September 1993, 
we noticed such a CS component in the spectrum of HD 204965 
and in October 1995 this component had disappeared.

Several of our programme stars display the sodium doublet. These 
stars are listed in Table 5, which includes some of their spectroscopic 
characteristics and the extension of the circumstellar shell, for which 
the relation of Higurashi and Hirata (1978) was used, as in the case of 
CaII-K. Five of the stars in Table 5 possess a CS component on the 
CaII-K line.

The 14 other stars observed in the NaID region do not display the NaID 
doublet, either because it is too weak or because it is completely masked 
by the telluric lines which overcrowd this spectral domain.

A comparison of the RVs of the CaII-K CS component (Table 4) with 
the NaI doublet (Table 5) reveals that they are always similar to each 
other. It may be assumed that these ions are associated in the 
expansion/contraction movement of the circumstellar shell. It is also 
noted that the absorption shell of NaI is inside that of CaII.
Special mention is made of the sodium doublet of HD 225180, 
shown in Figure 4. These lines are strong and all the lines of the 
optical spectrum possess a shell component. The NaI doublet presents 
a double structure in which a photospheric and a red-shift shell 
component may be identified. The shell component has an asymmetric 
background.

The RVs of the points of inflexion on each blue edge are -39.5 $km\ s^{-1}$
1 (5890 \AA) and -36.0 $km\ s^{-1}$  (5896 \AA) respectively. Close to 
the stellar continuum, at $I = 0.96$, the total width of each line is 83.2 
$km\ s^{-1}$ and 77.7 $km\ s^{-1}$  and the RV line at this level is -21.0 and -20.7 
$km\ s^{-1}$.

Finally, the velocity gradient along each line is +12.6 $km\ s^{-1}$ and 
+12.1 $km\ s^{-1}$ respectively, indicating a circumstellar movement 
towards the star (onfall).

\section{Discussion}
\subsection{Difference of $V_r$ between $H_8$ and CaII-K lines}
As noticed in Paper I, we may remark in some cases a difference 
between the radial velocity deduced from $H_8$ and that from CaII-K. 
Our sample is now sufficiently large to see clearly from the data of 
table 3 that this is not a general property since nearly half of the 
sample does not exhibit such a gradient, including $\lambda \ Boo$ itself. 
Moreover, no correlation has been found between the presence or 
absence of this gradient and one of the criteria, or suspected criteria, 
such as the IR excess or UV depression. However, as can be seen from 
Table 1, nearly all the stars studied are spectroscopic binaries or have 
a variable radial velocity. Thus we can assume that this difference is 
due to the binarity of these stars.

\subsection{Variation of $V_r$ between two periods of observations}
A variation of $V_r$ between two or more periods of observations could 
be an indication of binarity or of pulsation. There are too few 
measurements for a clear conclusion to be reached from our 
observations. We can only remark that of the six stars for which we 
have data in the same spectral regions at two or three epochs only 
one (HD 204965) exhibits a severe change of radial velocity in one 
year and one (HD 38545) a moderate change, while the other four (HD 
31295, HD 34787, HD 210418 and HD 220061) are seen to undergo no, 
or only a small, change.

\subsection{Difference of $V_r$ between the photospheric lines and the circumstellar lines}
We can consider the difference of $V_r$ between the photospheric line 
and the circumstellar line for stars for which a circumstellar 
component has been detected on the CaII-K line. This difference may 
be interpreted as being the radial component of the circumstellar gas 
movement. This gradient is positive for the following stars, indicating 
a fall of the gas on the star:  HD 2904, HD 16811. For the following 
stars the gradient is negative, indicating an expansion of the gas: 
HD 5789, HD 34787 (in 1994 and 1995), HD 141851, HD 161868, 
HD 217782. The gradient is slightly negative for HD 38545. Two stars 
of the 1993 mission that also show an onfall, HD 16955 and HD 
204965, can be added to the first of the above-mentioned group.
\subsection{Rotation velocities of $\lambda \ Bootis$ stars with a CS component}
Holweger and Rentzsch-Holm (1995) have shown that the $\lambda \ Bootis$ 
stars with detectable CS gas are to be found preferably among rapid 
rotators. This is still true if we consider our results. With the 
exception of HD 204965 and HD 221756, all the stars of Table 4 
possess $Vsin\ i$ values that are greater than 100 $km\ s^{-1}$. However, on the 
basis of their photometric data, these two stars cannot be considered 
to be $\lambda \ Bootis$ stars. It may also be remarked that not all $\lambda \ Bootis$ stars 
with a $Vsin\ i$ value in excess of 100 $km\ s^{-1}$ exhibit a CS component.

Holweger and Rentzsch-Holm thus conclude that $Vsin\ i$ is the prime 
factor responsible for the presence or absence of CS absorption in 
CaII-K. They also make the assumption that $\lambda \ Bootis$ stars are pre-
main-sequence objects that are still accreting material left over from 
the protostellar phase. Our discussion in 5.3 shows that it is difficult to 
confirm or rule out this assumption on the basis of the difference of 
Vr between the photospheric lines and the circumstellar lines.

\section{Conclusions}
Our observations confirm the presence of CS gas around eight $\lambda \ Bootis$ 
stars, more particularly among the most rapid rotators. This is in line 
with the recent study of Holweger and Rentzsch-Holm (1995), who 
conclude that these stars are in the pre-main-sequence phase of 
evolution. Furthermore, our data show that the intensity of the CS component is variable on a timescale of months or years. As 
the actual time resolution of the observations does not cover a timescale 
of days we cannot arrive at a conclusion for this timescale. The analysis 
of the photometric data confirms that four $\lambda Bootis$ candidates can be 
excluded from the $\lambda Bootis$ group.

\acknowledgements{This research has made use of the SIMBAD 
database, operated at the CDS, Strasbourg, France. One of us (BH) is 
grateful to the Fonds National suisse de la Recherche Scientifique, 
which gave its support to part of the research.}

\newpage

{\small
\begin{table*}
\scriptsize
\caption{General data for the observed stars}
\begin{flushleft}
\begin{tabular}{r r r r r r r r r r r r r r r}
\hline
name			&HR	&HD	&S1		&V	&Sp			&Sp		&S2	&RV	&$v sin\ i$	&B2-Vl	&d	&$\Delta m_2$	&$\Delta$	&$T_{\mathrm eff}$\\
			&	&	&&&Abt \& Morrell	&\\
\hline
			&129	&2904	&3		&6.42	&A0Vnn($\lambda \ Boo$)	&A0Vn		&2	&-10	&225:		&-0.149	&1.480	&		&0.461		&9400\\
			&283	&5789	&3		&6.04	&B9.5Vnn($\lambda \ Boo$)	&B9.5Vn		&2	&+lSB	&230:&\\
$\gamma ^1$ Ari		&545	&11503	&3		&4.83	&A0Vp($\lambda \ Boo$)n	&B9.5IVn		&1	&+4V	&185&\\
$\mu$ Ari		&793	&16811	&2		&5.73	&A0Vn			&A0IVn		&1	&-7V	&160		&-0.161	&1.446	&		&0.444		&9500\\
			&1137	&23258	&3		&6.10	&A0Vp($\lambda \ Boo$)	&A0V		&2	&+15SB	&110		&-0.141	&1.381	&		&0.494		&9300\\
$\pi ^2$ Ori		&1544	&30739	&3		&4.36	&A0Vp($\lambda \ Boo$)n	&A0.5IVn	&1	&+24SB	&195		&-0.136	&1.542	&		&0.539		&9200\\
$\pi ^1$ Ori		&1570	&31295	&1		&4.65	&A0Vp($\lambda \ Boo$)	&A0Va$\lambda \ Boo$	&1	&+13V	&105		&-0.091	&1.372	&-0.010		&0.524		&8800\\
16 Cam			&1751	&34787	&3		&5.23	&B9.5Vp($\lambda \ Boo$)n	&A0Vn		&2	&+12SB	&200:		&-0.163	&1.534	&		&0.458		&9400\\
			&1853	&36496	&4		&6.26	&A5Vn			&A8Vn		&2	&-15V?	&180		& 0.034	&1.196	&-0.029		&0.354		&7700\\
131 Tau			&1989	&38545	&1		&5.72	&A2IVn+sh		&A2Va$\lambda \ Boo$	&1	&+21V	&175		&-0.087	&1.526	&-0.015		&0.588		&8700\\
$\xi$ Aur		&2029	&39283	&4		&4.99	&A1IVp			&A1Va		&1	&-12V?	&55		&-0.115	&1.492	&		&0.598		&9000\\
53 Aur			&2425	&47152	&3		&5.79	&A2Vp($\lambda \ Boo$)	&B9np		&2	&+18V?	&25		&-0.161	&1.368	&		&0.413		&9500\\
14 Com			&4733	&108283	&3		&4.95	&A9Vp($\lambda \ Boo$)	&A9IVnpSr	&1	&-4SB1	&85 		& 0.081	&1.419	&0.018		&0.539		&7400\\
9 CVn			&4811	&109980	&3		&6.37	&A6Vp($\lambda \ Boo$)	&A7Vn		&2	&-15	&255:		 &0.006	&1.272	&-0.023		&0.445		&7900\\
27 Vir			&4824	&110377	&3		&6.19	&A6Vp($\lambda \ Boo$)	&A7Vn		&2	&+9SB	&160		 &0.010 &1.266	&-0.038		&0.401		&7900\\
$\rho$ Vir		&4828	&110411	&1		&4.88	&A0Vp			&A0Va($\lambda \ Boo$)	&1	&+2SB	&140		&-0.092	&1.407	&-0.021		&0.525		&8800\\
			&4875	&111604	&2		&5.89	&A5Vp($\lambda \ Boo$)	&A3V		&2	&-14V	&180		&-0.007	&1.415	&-0.037		&0.482		&8000\\
41 Vir			&4900	&112097	&3		&6.25	&F0Vp($\lambda \ Boo$)	&B9III		&2	&-l0SB	&61		& 0.080	&1.127	&-0.020		&0.283		&7400\\
25 CVn			&5127	&118623	&3		&4.82	&F0Vp($\lambda \ Boo$)n	&A7Vn		&1	&-6V1	&90		& 0.040	&1.280	&-0.017		&0.420		&7600\\
$\lambda \ Boo$		&5351	&125162	&1		&4.18	&A0Vp($\lambda \ Boo$)	&A0Va$\lambda \ Boo$	&1	&-8	&110		&-0.081	&1.400	&-0.013		&0.546		&8700\\
			&5368	&125489	&3		&6.19	&F0Vp($\lambda \ Boo$)	&A7V		&1	&-22V	&145&\\
36 Ser			&5895	&141851	&2		&5.11	&A3Vp(*)n		&A3Vn		&2	&-8V	&185		&-0.049	&1.388	&-0.041		&0.491		&8400\\
$\epsilon$ Her		&6324	&153808	&3		&3.92	&A0IVp($\lambda \ Boo$)	&A0IV		&1	&-25SBO	&50		&-0.161	&1.394	&		&0.409		&9500\\
			&6532	&159082	&3		&6.42	&A0IVp($\lambda \ Boo$)	&B9.5V		&2	&-12SBO	&30&\\
$\gamma$ Oph		&6629	&161868	&2		&3.75	&A0Vp(*)n		&A0Van		&1	&-7SB?	&185		&-0.124	&1.459	&		&0.549		&9000\\
			&7903	&196821	&3		&6.08	&A0IIIp($\lambda \ Boo$)s	&A0III		&2	&-37SB?	&10		&-0.185	&1.522	&		&0.426		&10200\\
 			&8237	&204965	&2		&6.02	&A2Vp(*)		&A3V		&2	&-17SB	&85		&-0.090	&1.606	 &0.005		&0.672		&8700\\
$\theta$ Peg		&8450	&210418	&2		&3.53	&A2V			&A2m1IV-V	&1	&-6SB2	&130		&-0.086	&1.461	&-0.013		&0.564		&8700\\
9 Lac			&8613	&214454	&3		&4.63	&F0Vp($\lambda \ Boo$)	&A7IV-V		&1	&+12SB	&93		& 0.050	&1.298	&-0.003		&0.453		&7600\\
2 And			&8766	&217782	&4		&5.10	&A1V			&A3Vn		&2	&+2SB	&195		&-0.075	&1.567	&-0.023		&0.605		&8600\\
$\tau$ Peg		&8880	&220061	&2		&4.60	&A5Vp($\lambda \ Boo$)	&A5V		&1	&+16V	&135		&-0.005	&1.382	&-0.019		&0.507		&8000\\
15 And			&8947	&221756	&1		&5.59	&A1Vp(*)		&A1Va($\lambda \ Boo$)	&1	&+13V	&75		&-0.089	&1.459	&-0.002		&0.595		&8700\\
9 Cas			&9100	&225180	&3		&5.88	&A1Vp($\lambda \ Boo$)	&A1II-III	&1	&-18V	&25		&0.259	&1.805	&		&0.798\\
\hline
\multicolumn{15}{l}{Notes to table} \\
\multicolumn{15}{l}{Sources of $\lambda \ Boo$ type, column 4: 1. Gray (1988) 2. Abt (1984) 3. Abt and Morrell (1995) 4. Andrillat et al. (1995)}\\
\multicolumn{15}{l}{Sources of spectral types, column 8: 1. Gray and Garrison (1987, 1989a,b) 2. Cowley et al. (1969)}\\
\multicolumn{15}{l}{$^*$ in column 6 is for 4481 wk}\\
\hline
\end{tabular}
\end{flushleft}
\end{table*}
}

\begin{table*}
\caption{Log of observations (1994-1995)}
\begin{flushleft}
\begin{tabular}{r r r r r r r}
\hline
star	&line(s)	&date&HJD-2400000&	exposure  &	resolution&S/N\\HD&&d/m/y		&&time (min.)\\
\hline
2904&$H_8$-CaIIK&	01/11/95&	50023.480&	158&	14000&	180\\
5789	&$H_8$-CaIIK&	31/10/95&	50022.417&	105&	14000&	450\\	
	&NaID		&02/11/95&	50024.469&	120&	21000&	320\\
11503	&$H_8$-CaIIK&	10/11/94&	49667.420&	43&	23000&	380\\	
	&$H_\gamma$&	11/11/94&	49668.389&	47&	25000&	470\\
16811	&$H_8$-CaIIK&	30/10/95&	50021.474&	76&	14000&	390\\	
	&NaID		&03/11/95&	50025.419&	132&	21000&	420\\
23258	&$H_8$-CaIIK&	01/11/95&	50022.585&	98&	14000&	300\\	
	&NaID		&03/11/95&	50025.551&	164&	21000&	300\\
30739	&$H_\gamma$&	11/11/94&	49668.436&	66&	25000&	430\\
31295	&$H_\gamma$&	11/11/94&	49668.510&	66&	25000&	450\\
34787	&$H_8$-CaIIK&	01/11/94&	49667.563&	62&	23000&	350\\	
        &$H_8$-CaIIK&	02/11/95&	50023.675&	108&	14000&	280\\	
        &$H_\gamma$&	12/11/94&	49668.556&	62&	25000&	420\\	
	&NaID		&03/11/95&	50024.559&	65&	21000&	460\\
36496	&$H_8$-CaIIK&	31/10/95&	50021.632&	120&	14000&	270\\
38545	&$H_8$-CaIIK&	11/11/94&	49667.630&	113&	23000&	350\\	
	&$H_\gamma$&	12/11/94&	49668.653&	102&	25000&	430\\
	&NaID		&04/11/95&	50025.662&	150&	21000&	300\\
39283	&$H_8$-CaIIK&	31/10/95&	50021.694&	43&	14000&	210\\	
	&$H_{\beta}$*&	05/11/95&	50026.677&	101&	17000&	600\\	
	&NaID		&03/11/95&	50024.606&	64&	21000&	480\\
47152	&$H_8$-CaIIK&	01/11/95&	50022.656&	89&	14000&	160\\	
	&NaID		&03/11/95&	50024.673&	120&	21000&	150\\
108283	&$H_8$-CaIIK&	05/05/95&	49843.363&	96&	23000&	350\\	
	&NaID*		&08/05/95&	49846.356&	60&	34000&	400\\
109980	&NaID		&10/05/95&	49848.389&	167&	34000&	420\\
110377	&NaID		&09/05/95&	49847.383&	167&	34000&	470\\
110411	&$H_8$-CaIIK&	06/05/95&	49844.382&	111&	23000&	450\\	
	&$H_\gamma$&	07/02/95&	49755.582&	19&	15000&	430\\	
	&$H_{\beta}$&	10/02/95&	49758.608&	38&	17000&	500\\	
	&NaID		&08/05/95&	49846.404&	59&	34000&	520\\	
	&$H_\alpha$	&10/02/95&	49758.671&	23&	23000&	380\\
111604	&$H_8$-CaIIK&	07/05/95&	49845.395&	161&	23000&	380\\	
	&$H_\gamma$&	07/02/95&	49755.614&	46&	15000&	420\\
112097	&$H_\gamma$&	07/02/95&	49755.658&	66&	15000&	260\\
118623	&$H_8$-CaIIK&	05/05/95&	49843.438&	77&	23000&	400\\	
	&$H_\gamma$&	07/02/95&	49755.688&	16&	15000&	470\\
	&NaID		&08/05/95&	49846.448&	58&	34000&	450\\	
	&$H_\alpha$	&10/02/95&	49758.690&	27&	23000&	420\\
125162	&$H_8$-CaIIK&	05/05/95&	49843.490&	40&	23000&	450\\	
	&NaID		&08/05/95&	49846.484&	23&	34000&	480\\
125489	&$H_8$-CaIIK*&	06/05/95&	49844.995&	300&	23000&	370\\	
	&NaID*		&09/05/95&	49847.983&	187&	34000&	360\\
141851	&$H_8$-CaIIK&	06/05/95&	49843.547&	77&	23000&	260\\	
	&NaID		&08/05/95&	49846.530&	91&	34000&	480\\
153808	&$H_8$-CaIIK&	06/05/95&	49843.584&	22&	23000&	380\\	
	&NaID		&09/05/95&	49846.568&	15&	34000&	450\\
159082	&$H_8$-CaIIK*&	07/05/95&	49845.089&	225&	23000&	300\\
	&NaID		&11/05/95&	49848.564&	103&	34000&	901\\
161868	&$H_8$-CaIIK&	06/05/95&	49843.607&	24&	23000&	310\\
	&NaID		&09/05/95&	49846.587&	22&	34000&	530\\
196821	&$H_8$-CaIIK&	31/10/95&	50022.297&	111&	14000&	200\\
	&NaID		&02/11/95&	50024.305&	120&	21000&	220\\
204965	&$H_8$-CaIIK&	30/10/95&	50021.328&	114&	14000&	300\\
210418	&$H_8$-CaIIK&	10/11/94&	49667.253&	24&	23000&	300\\	
	&$H_8$-CaIIK&	01/11/95&	50023.277&	26&	14000&	400\\	
	&$H_\gamma$&	11/11/94&	49668.266&	51&	25000&	360\\
\hline
\end{tabular}
\end{flushleft}
\end{table*}
\newpage
\begin{table*}
\hspace{0cm}{\bf Table 2 (continued).} 
\begin{flushleft}
\begin{tabular}{r r r r r r r}
\hline
star	&line(s)	&date&HJD-2400000&	exposure  &	resolution&S/N\\
HD&&d/m/y		&&time (min.)\\
\hline
214454	&$H_8$-CaIIK&	10/11/94&	49667.290&	64&	23000&	300\\	
	&$H_\gamma$&	11/11/94&	49668.307&	54&	25000&	300\\
217782	&$H_8$-CaIIK&	31/10/95&	50022.359&	48&	14000&	380\\	
	&NaID		&03/11/95&	50025.341&	82&	21000&	450\\
220061	&$H_8$-CaIIK&	10/11/94&	49667.357&	102&	23000&	250\\	
	&$H_8$-CaIIK&	01/11/95&	50023.310&	55&	14000&	380\\	
	&$H_\gamma$&	11/11/94&	49668.349&	51&	25000&	300\\
221756	&$H_8$-CaIIK&	01/11/95&	50023.373&	120&	14000&	400\\
	&NaID		&02/11/95&	50024.388&	100&	21000&	520\\
225180	&$H_8$-CaIIK&	30/10/95&	50021.407&	98&	14000&	300\\
	&NaID&		04/11/95&	50026.354&	300&	21000&	400\\
\hline
\multicolumn{7}{r}{*...mean of two spectra (HJDs are averaged, and exposure times added)}\\
\hline
\end{tabular}
\end{flushleft}
\end{table*}

\begin{table*}
\begin{flushleft}
\caption{Radial velocities ($km\ s^{-1}$) of photospheric lines (The RV value is taken from the BS Catalogue)}
\begin{tabular}{l l r r r}
\multicolumn{5}{l}{1) NOVEMBER 1994}\\
&&		$H_8$&	CaIIK&	$H_{\gamma}$\\
HD 11503&	parabolic fit	&+8.4	&+17.3&	+12.0\\
RV = +4 V&	gravity centre	&+8.7	&+16.8	&+12.0	\\
	&	mean  	&+8.6	&+17.1	&+12.0	\\
	&	\multicolumn{4}{l}{mean of the 6 measurements =+12.5 $\pm$ 3.5}\\							
HD 30739&	parabolic fit			&&&+28.7\\
RV = +24 SB&	gravity centre			&&&+28.7	\\
	&	mean			&&&+28.7	\\
	&	\multicolumn{4}{l}{mean of the 2 measurements = +28.7	}		\\
HD 31295&	parabolic fit 			&&&+13.7\\
RV = +13 V&	gravity centre 			&&&+13.5	\\
	&	mean 			&&&+13.6	\\
	&	\multicolumn{4}{l}{mean of the 2 measurements = +13.6	}		\\
HD 34787	&parabolic fit   	&+8.3	&+10.1	&+8.7\\
RV = +12 SB&	gravity centre   	&+8.1	&+14.5	&+9.0	\\
		&mean   	&+8.2	&+12.3	&+8.9	\\
		&\multicolumn{4}{l}{mean of the 6 measurements = +9.8 $\pm$ 2.2}\\			
HD 38545	&parabolic fit   	&+11.5	&+11.0	&+12.1\\
RV = +21 V	&gravity centre   	&+11.4	&+12.6	&+12.9	\\
		&mean   	&+11.5	&+11.8	&+12.5	\\
		&\multicolumn{4}{l}{mean of the 6 measurements = +11.9 $\pm$ 0.7}	\\
HD 210418	&parabolic fit   	&-8.9	&-6.7	&-9.7\\
RV = -6 SB2	&gravity centre   	&-8.4	&-10.1	&-10.1	\\
		&mean   	&-8.7	&-8.4	&-9.9	\\
		&\multicolumn{4}{l}{mean of the 6 measurements = -9.0 $\pm$ l.2}	\\
HD 214454	&parabolic fit   	&+11.2	&+9.2	&+10.5\\
RV = +12 SB	&gravity centre	&+11.3	&+9.1	&+11.2	\\
		&mean 	&+11.2	&+9.1	&+10.8	\\
		&\multicolumn{4}{l}{mean of the 6 measurements = +10.4 $\pm$ 0.9}	\\
HD 220061	&parabolic fit  	&+0.8	&+9.2 	&+3.9\\
RV = +16 V	&gravity centre 	&+2.1 	&+8.6 	&+3.4	\\
		&mean   	&+1.5	&+8.9	&+3.7	\\
		&\multicolumn{4}{l}{mean of the 6 measurements = +4.5 $\pm$ 3.1}	\\
\multicolumn{5}{l}{2) FEBRUARY 1995}\\
&&		$H_{\gamma}$	&$H_{\beta}$&	$H_{\alpha}$\\
HD 10411	&parabolic fit	&-9.2	&-7.7	&-8.2\\
RV = +2 SB	&gravity centre	&-9.2	&-8.4	&-8.8	\\
		&mean	&-9.2	&-8.1	&-8.5	\\
		&\multicolumn{4}{l}{mean of the 6 measurements = -8.6 $\pm$ 0.5}	\\
HD 111604	&parabolic fit	&-12.2		\\
RV =-14 V	&gravity centre	&-17.6			\\
		&mean	&-14.9		\\
		&\multicolumn{4}{l}{mean of the 2 measurements = -14.9	}		\\
HD 112097	&parabolic fit	&-7.5			\\
RV = -l0 SB	&gravity centre	&-7.9			\\
		&mean	&-7.7					\\
		&\multicolumn{4}{l}{mean of the 2 measurements= -7.7	}		\\
HD 118623	&parabolic fit	&-7.1		&&-4.3\\
RV = -6 V	&gravity centre	&-7.2		&&-5.3	\\
		&mean	&-7.2		&&-4.8	\\
		&\multicolumn{4}{l}{mean of the 4 measurements = -6.0 $\pm$ 1.2	}\\		
\multicolumn{5}{l}{3) MAY 1995}\\
&&		$H_8$	&CaIIK\\
HD 108283	&parabolic fit	&-21.5	&2.3\\
RV = -4 SB	&gravity centre	&-22.3&	2.7	\\
		&mean	&-21.9	&2.5	\\
		&\multicolumn{4}{l}{mean of the 4 measurements = -9.7 $\pm$ 12.2	}\\	
HD 110411	&parabolic fit 	&-9.4	&-6.8\\
RV = +2 SB	&gravity centre	&-9.8	&-6.9	\\
		&mean 	&-9.6	&-6.8	\\
		&\multicolumn{4}{l}{mean of the 4 measurements = -8.2 $\pm$ 1.4	}\\	
\end{tabular}
\end{flushleft}
\end{table*}

\begin{table*}
\hspace{0cm}{\bf Table 3 (continued).} 
\begin{flushleft}
\begin{tabular}{l l r r r}
HD 111604	&parabolic fit	&-22.1	&-10.2\\
RV = -14 V	&gravity centre	&-22.4	&-9.7	\\
		&mean	&-22.3	&-10.0	\\
		&\multicolumn{4}{l}{mean of the 4 measurements = -16.1 $\pm$ 6.2}\\
HD 118623	&parabolic fit	&-16.5	&-8.9\\
RV = -6 V	&gravity centre	&-16.7	&-9.0	\\
		&mean	&-16.6	&-9.0	\\
		&\multicolumn{4}{l}{mean of the 4 measurements = -12.8 $\pm$ 3.8}\\		
HD 125162	&parabolic fit	&-9.1	&-8.9\\
RV = -8		&gravity centre	&-9.5	&-8.2	\\
		&mean	&-9.3	&-8.6	\\
		&\multicolumn{4}{l}{mean of the 4 measurements = -8.9 $\pm$ 0.4	}\\	
HD 125489	&parabolic fit	&-30.1	&-21.9\\
RV = -22 V	&gravity centre	&-30.7	&-22.3\\
		&mean	&-30.4	&-22.1	\\
		&\multicolumn{4}{l}{mean of the 4 measurements = -26.2 $\pm$ 4.2}\\		
HD 141851	&parabolic fit	&-18.7	&-13.3\\
RV = -8 V	&gravity centre	&-18.2	&-10.2	\\
		&mean	&-18.5	&-11.8	\\
		&\multicolumn{4}{l}{mean of the 4 measurements = -15.1 $\pm$ 3.5}\\		
HD153808	&parabolic fit	&-27.1	&-29.3\\
RV = -25SBO	&gravity centre	&-27.4	&-29.1	\\
		&mean	&-27.2	&-29.2	\\
		&\multicolumn{4}{l}{mean of the 4 measurements = -28.2 $\pm$ 1.0}\\		
HD 159082	&parabolic fit	&-59.8	&-57.7	\\
RV = -12 SBO	&gravity centre	&-58.7	&-58.5	\\
		&mean	&-59.2	&-58.1	\\
		&\multicolumn{4}{l}{mean of the 4 measurements = -58.7 $\pm$ 0.72}\\		
HD 161868	&parabolic fit	&-20.5	&-20.2\\
RV = -7 SB	&gravity centre	&-20.4	&-19.9	\\
		&mean	&-20.5	&-20.0	\\
		&\multicolumn{4}{l}{mean of the 4 measurements = -20.3 $\pm$ 0.2}\\					\multicolumn{5}{l}{4) OCTOBER-NOVEMBER 1995}\\							
&&$H_8$	&CaIIK&	$H_{\beta}$	\\
HD 2904 	&parabolic fit	&-23.2	&-3.9	\\
RV = -10	&gravity centre	&-25.6	&-4.3		\\
		&mean	&-24.4	&-4.1		\\
		&\multicolumn{4}{l}{mean of the 4 measurements = -14.2 $\pm$ 10.2}\\			
HD 5789		&parabolic fit	&1.1&	16.8	\\
RV = +l SB	&gravity centre&	0.8&	17.2		\\
		&mean	&0.9	&17.0				\\
		&\multicolumn{4}{l}{mean of the 4 measurements = 9.00 $\pm$ 8.1}		\\	
HD 16811	&parabolic fit	&-1.3	&5.3	\\
RV = -7 V	&gravity centre	&-1.1	&6.1		\\
		&mean	&-1.2&	5.7		\\
		&\multicolumn{4}{l}{mean of the 4 measurements = 2.2 $\pm$ 3.5}		\\	
HD 23258	&parabolic fit	&14.5&	16.4	\\
RV = +15 SB	&gravity centre	&14.6&	16.4		\\
		&mean 	&14.6	&16.4		\\
		&\multicolumn{4}{l}{mean of the 4 measurements = 15.5 $\pm$ 0.9}\\			
HD 34787	&parabolic fit	&9.2&	10.2	\\
RV = +12 SB	&gravity centre&	8.7	&13.3		\\
		&mean	&9.0	&11.8		\\
		&\multicolumn{4}{l}{mean of the 4 measurements = 10.4 $\pm$ 1.8}\\			
HD 36496	&parabolic fit	&-24.8	&-13.2	\\
RV = -15 V?	&gravity centre	&-25.1	&-13.3	\\	
		&mean	&-25.0	&-13.2		\\
		&\multicolumn{4}{l}{mean of the 4 measurements = -19.1 $\pm$ 5.9}\\			
HD 39283	&parabolic fit	&-19.1	&-17.1	&-17.4\\
RV = -12 V?	&gravity centre	&-18.9	&-17.2	&-17.4\\	
		&mean	&-19.0	&-17.2	&-17.4	\\
		&\multicolumn{4}{l}{mean of the 6 measurements = -17.9 $\pm$ 0.8}	\\
\end{tabular}
\end{flushleft}
\end{table*}

\begin{table*}
\hspace{0cm}{\bf Table 3 (continued).} 
\begin{flushleft}
\begin{tabular}{l l r r r}
HD 47152	&parabolic fit	&14.5&	11.7	\\
RV = +18 V?	&gravity centre	&14.6&	11.8		\\
		&mean	&14.5	&11.8		\\
		&\multicolumn{4}{l}{mean of the 4 measurements = 13.1 $\pm$ 1.4}	\\
HD 196821	&parabolic fit	&-31.8	&-24.5	\\
RV = -37 SB?	&gravity centre	&-31.8	&-24.2		\\
		&mean	&-31.8	&-24.4		\\
		&\multicolumn{4}{l}{mean of the 4 measurements = -28.1 $\pm$ 3.73}	\\
HD 204965	&parabolic fit	&-16.1	&-14.4	\\
RV = -17 SB	&gravity centre	&-16.7	&-14.8		\\
		&mean	&-16.4	&-14.6		\\
		&\multicolumn{4}{l}{mean of the 4 measurements = -15.5 $\pm$ 0.9}	\\
HD 210418	&parabolic fit	&-9.1	&-5.4	\\
RV = -6 SB2	&gravity centre	&-8.9	&-4.8		\\
		&mean	&-9.0	&-5.1				\\
		&\multicolumn{4}{l}{mean of the 4 measurements = -7.1 $\pm$ 1.9}			\\
HD 217782	&parabolic fit	&-3.4&	3.2	\\
RV = +2 SB	&gravity centre	&-4.2&	2.6		\\
		&mean	&-3.8	&2.9		\\
		&\multicolumn{4}{l}{mean of the 4 measurements = -0.5 $\pm$ 3.36}\\
HD 220061	&parabolic fit	&2.7	&9.5	\\
RV = +16 V	&gravity centre	&3.4	&9.3		\\
		&mean	&3.1	&9.4		\\
		&\multicolumn{4}{l}{mean of the 4 measurements = 6.2 $\pm$ 3.2}\\			
HD 221756	&parabolic fit	&12.9	&14.2	\\
RV = +13 V	&gravity centre&	13.2	&14.5		\\
		&mean	&13.0	&14.3		\\
		&\multicolumn{4}{l}{mean of the 4 measurements = 13.7 $\pm$ 0.7}\\
HD 225180	&parabolic fit	&-21.9	&-14.0	\\
RV = - 18 V	&gravity centre	&-22.2	&-14.7		\\
		&mean	&-22.0	&-14.4		\\
		&\multicolumn{4}{l}{mean of the 4 measurements = -18.2 $\pm$ 3.8}\\			
\end{tabular}
\end{flushleft}
\end{table*}

\begin{table*}
\caption{Some data from the circumstellar component of CaII K.}
\begin{tabular}{|l| r |r |r| r |r|}
\hline
star HD&	date&	$V_r$ ($km\ s^{-1}$)&	$W_\lambda$(m\AA)&$\Delta V_{1/2}$&r/r*\\
\hline
2904		&01/11/95&	-2.7&	13.2&	13.35&	6.6\\
5789		&31/10/95&	-4.1&	14.1&	16.3&	5.8\\
16811&		30/10/95&	8.1&	19.7&	15.0&	4.8\\
16955$^{(1)}$	&20/09/93&	2.9	&9.6	&7.1&	8.0\\
34787		&11/11/94&	5.2	&18.1&	8.0	&8.5\\
34787		&02/11/95&	5.4	&16.0&	12.15&	6.5\\
38545		&11/11/94&	11.1	&72.2&	9.9	&6.8\\
141851&		06/05/95&	-27.5&	2.3&	6.9	&9.0\\
161868&		06/05/95&	-30.3&	6.0	&8.95&	7.5\\
204965$^{(2)}$&	20/09/93&	-18.2&	8.1	&6.9&	5.3\\
217782		&31/10/95&	-9.8&	20.0&	13.75&	5.9\\
\hline
\multicolumn{6}{|l|}{$^{(1)}$Star from Paper I (Vsini = 160 $km\ s^{-1}$)}\\
\multicolumn{6}{|l|}{$^{(2)}$Star from Paper I (Vsini = 85 $km\ s^{-1}$). The circumstellar component }\\
\multicolumn{6}{|l|}{has vanished on a spectrum dated 30/10/95.}\\
\multicolumn{6}{|l|}{$V_r$ = RV of the component}\\
\multicolumn{6}{|l|}{$\Delta V_{1/2}$ = width of the CaII K line at half-depth;}\\
\multicolumn{6}{|l|}{$W_\lambda$ (m\AA) = Equivalent width, reduced to the stellar continuum}\\
\multicolumn{6}{|l|}{r/r* = Extension of the circumstellar disk, in stellar radii (see text)}\\
\hline
\end{tabular}
\end{table*}

\begin{table*}
\caption{Some spectroscopic data from 8 programme stars which display the Na I D lines ($km\ s^{-1}$)}
\begin{tabular}{|l |r r r r r |r r r r r|r|}
\hline\multicolumn{1}{|r}{}
     &\multicolumn{5}{c}{Na I 5890}
     &\multicolumn{5}{c}{Na I 5896}
     &\multicolumn{1}{r|}{}\\
\hline
&$R_v$&	d & $\Delta V1_{1/2}$& R$V_{1/2}$&$W$&	$R_V$&d&$\Delta V_{1/2}$&		R$V_{1/2}$&	$W$&	r/r*\\
\hline
HD 5789&	-3.9&	0.821&	24.0&	-4.5&	95.5&		-3.7&	0.898&	21.0&	-4.6	&45.6&	4.7\\
HD 16811&	9.8&	0.764&	19.2&	10.6&	108.3&		11.0&	0.837&	21.3&	11.8&	73.7&	4.0\\
HD 34787&	6.5&	0.741&	20.3&	7.4&	131.4&		6.5&	0.821&	20.3&	6.8&	75.7&	4.6\\
HD 38545&	13.7&	0.836&	42.7&	17.5&	126.8&		17.1&	0.904&	-&	-&	74.8&	2.6\\
HD 108283&	0.3&	0.548&	43.8&	-0.8&	355.2&		2.8&	0.612&	46.5&	6.9&	398.9&	2.6\\
HD 159082&	-20.3&	0.628&	12.0&	-20.3&	92.4&		-20.4&	0.700&	12.0&	-20.4&	81.5&	1.8\\
HD 217782&	-7.4&	0.798&	24.1&	-8.4&	110.2&		-7.8&	0.883&	24.6&	-7.0&	59.2&	4.0\\
HD 225180&	-8.4&	0.359&	26.3&	-8.9&	433.9&		-8.6&	0.406&	25.2&	-9.2&	378.4&	1.0\\
\hline
\multicolumn{12}{|l|}{$R_{V}$ = radial velocity of the deepest point of the line}\\
\multicolumn{12}{|l|}{d = depth of the line, in unit of the continuum intensity (I = 1)}\\
\multicolumn{12}{|l|}{$\Delta V_{1/2}$ = width of the Na I D line at half-depth;}\\
\multicolumn{12}{|l|}{R$V_{1/2}$ = radial velocity of the line at half-depth}\\
\multicolumn{12}{|l|}{$W$ = equivalent width of the line (in m\AA)}\\
\multicolumn{12}{|l|}{r/r* = extension of the Na I circumstellar cloud, expressed in radii of the central star}\\
\multicolumn{12}{|l|}{The underlined stars are those which present a circumstellar absorption on the Ca II K photspheric }\\
\multicolumn{12}{|l|}{line (see Table 4).}\\
\hline
\end{tabular}
\end{table*}
\newpage

\begin{figure}
\epsfxsize=8.8cm
\leavevmode\epsffile{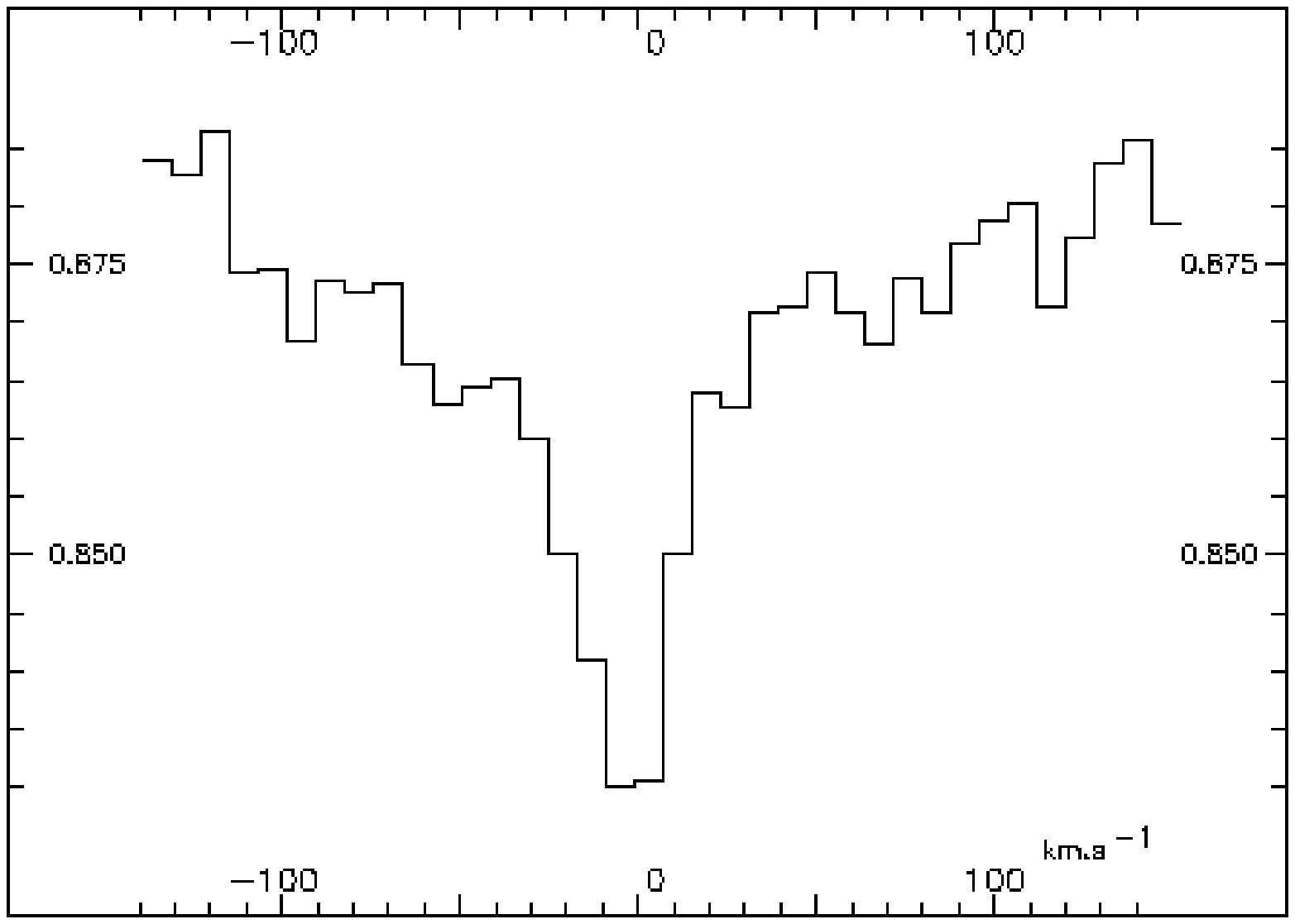}
\vspace{2mm}

{\bf Fig. 1a.} CaII-K line profile for HD 2904.
\end{figure}

\begin{figure}
\epsfxsize=8.8cm
\leavevmode\epsffile{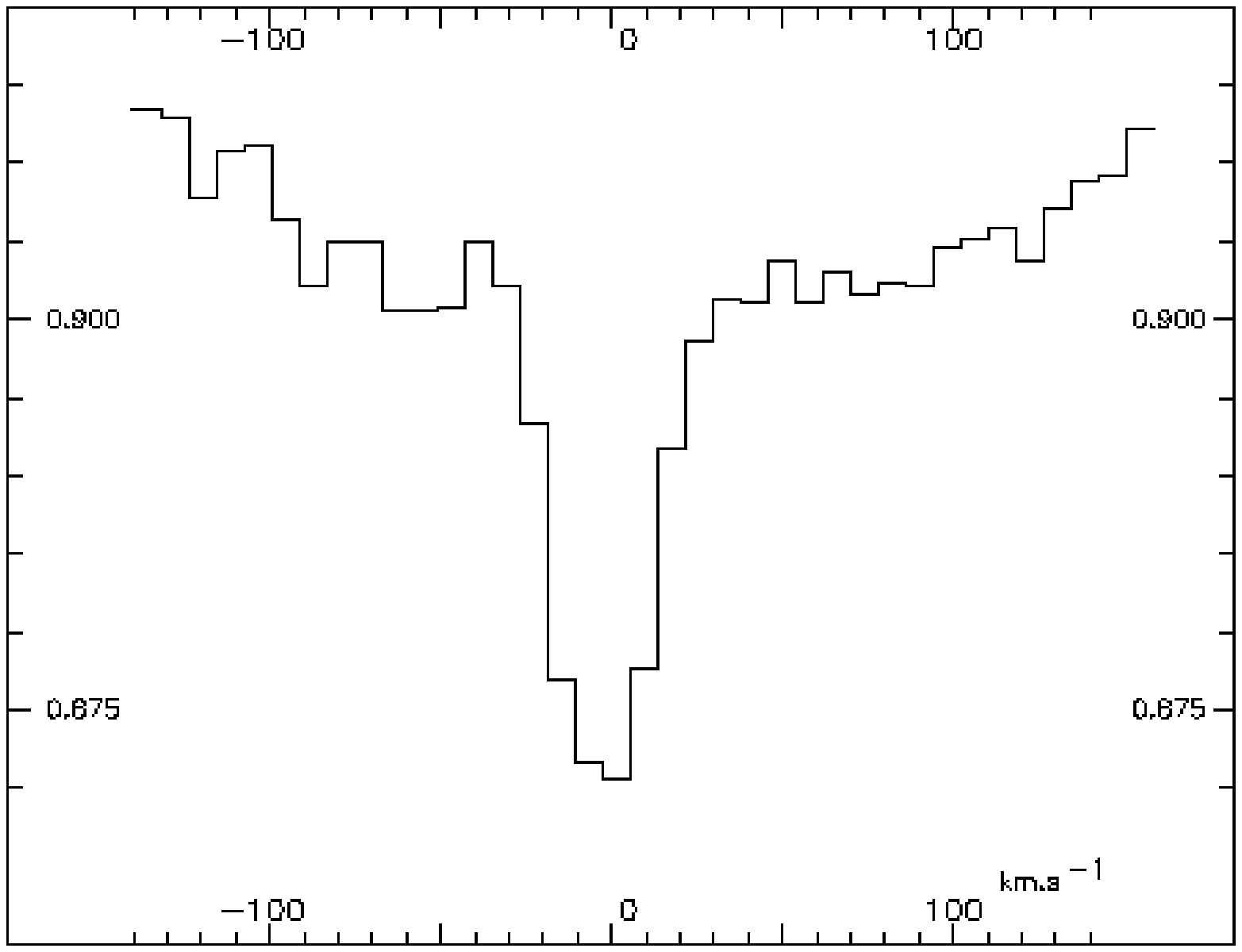}
\vspace{2mm}

{\bf Fig. 1b.} as Fig. 1a for HD 5789.
\end{figure}

\begin{figure}
\epsfxsize=8.8cm
\leavevmode\epsffile{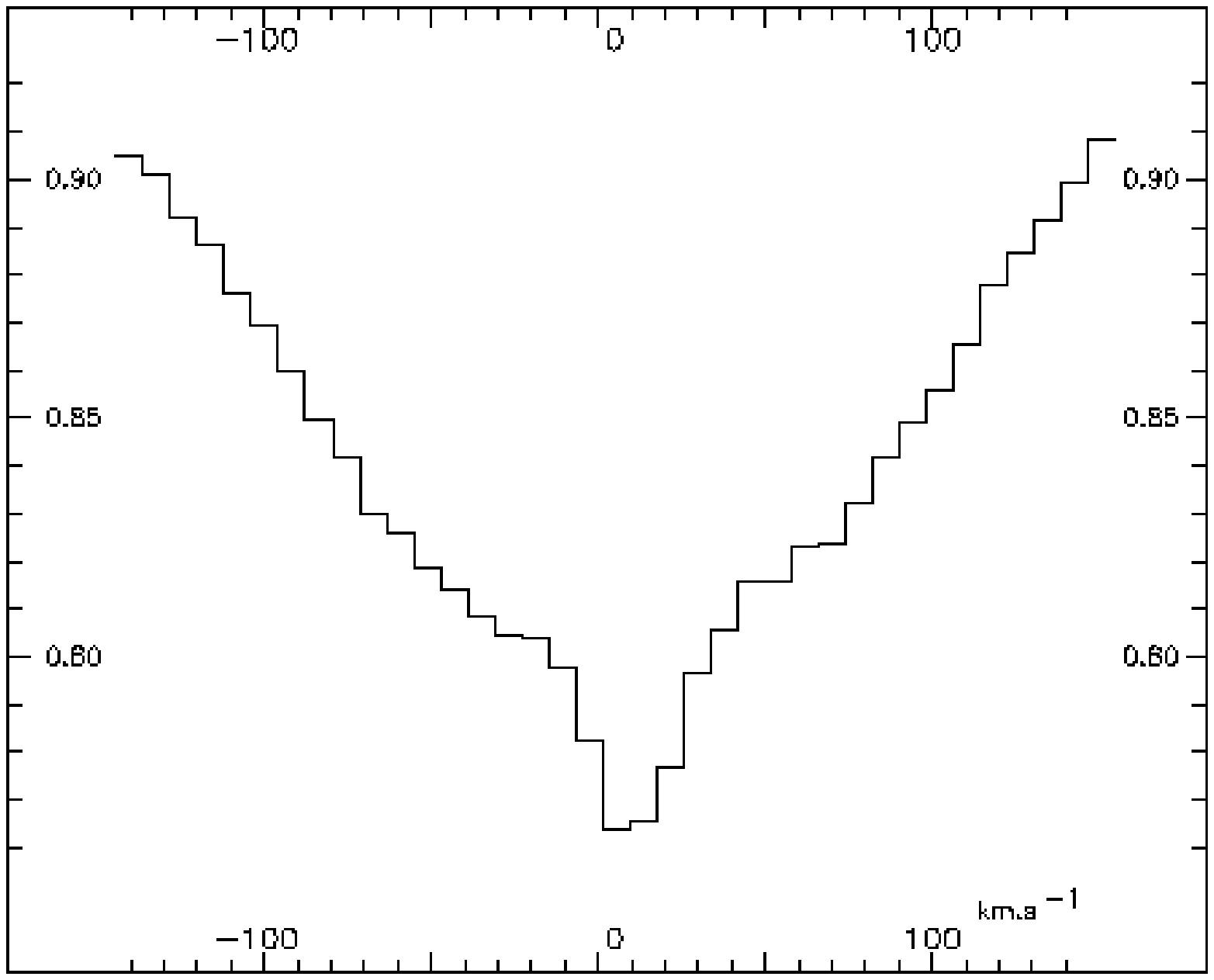}
\vspace{2mm}

{\bf Fig. 1c.} as Fig. 1a for HD 16811.
\end{figure}

\begin{figure}
\epsfxsize=8.8cm
\leavevmode\epsffile{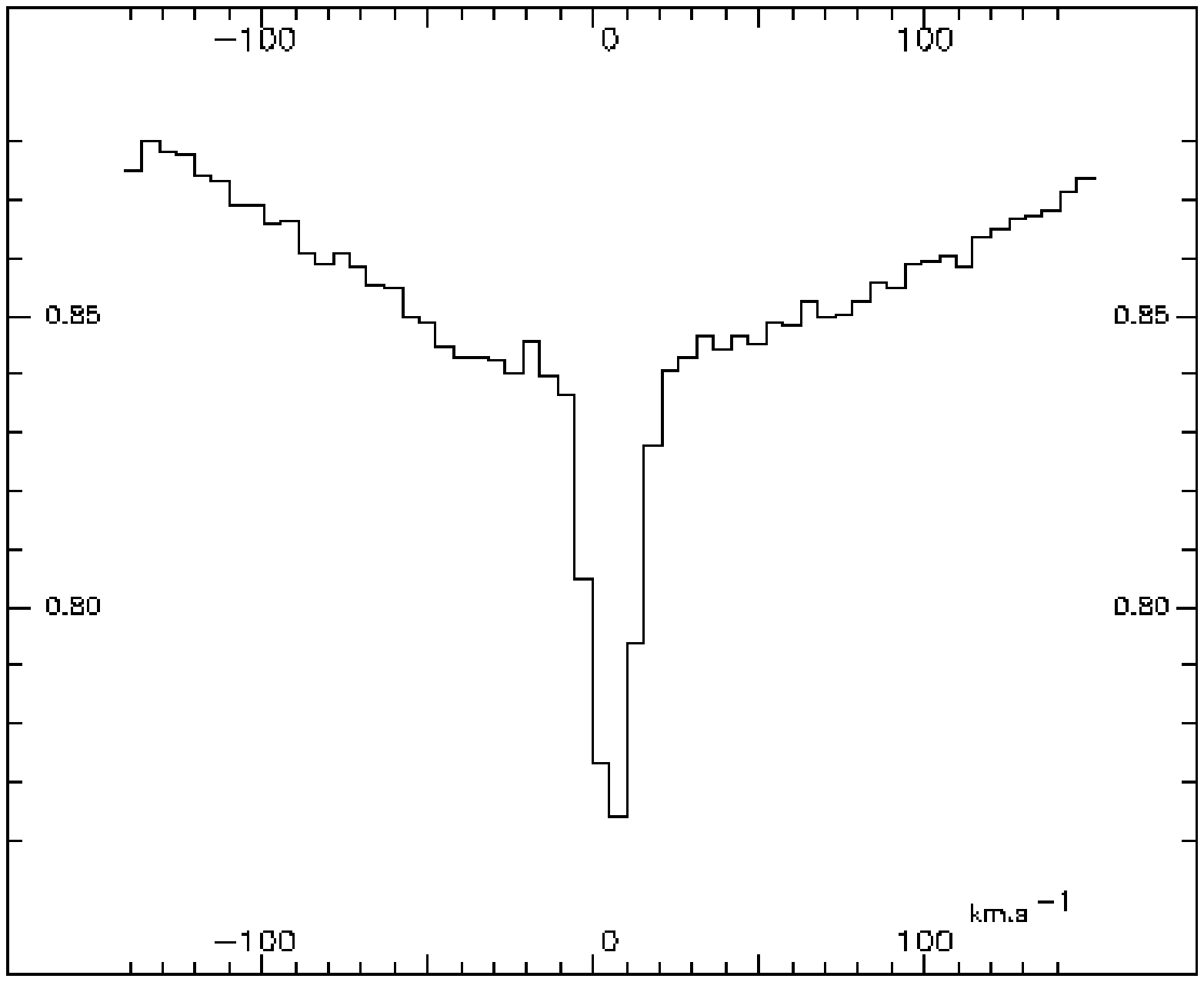}
\vspace{2mm}

{\bf Fig. 1d.} as Fig. 1a for HD 34787 11/11/94.
\end{figure}

\begin{figure}
\epsfxsize=8.8cm
\leavevmode\epsffile{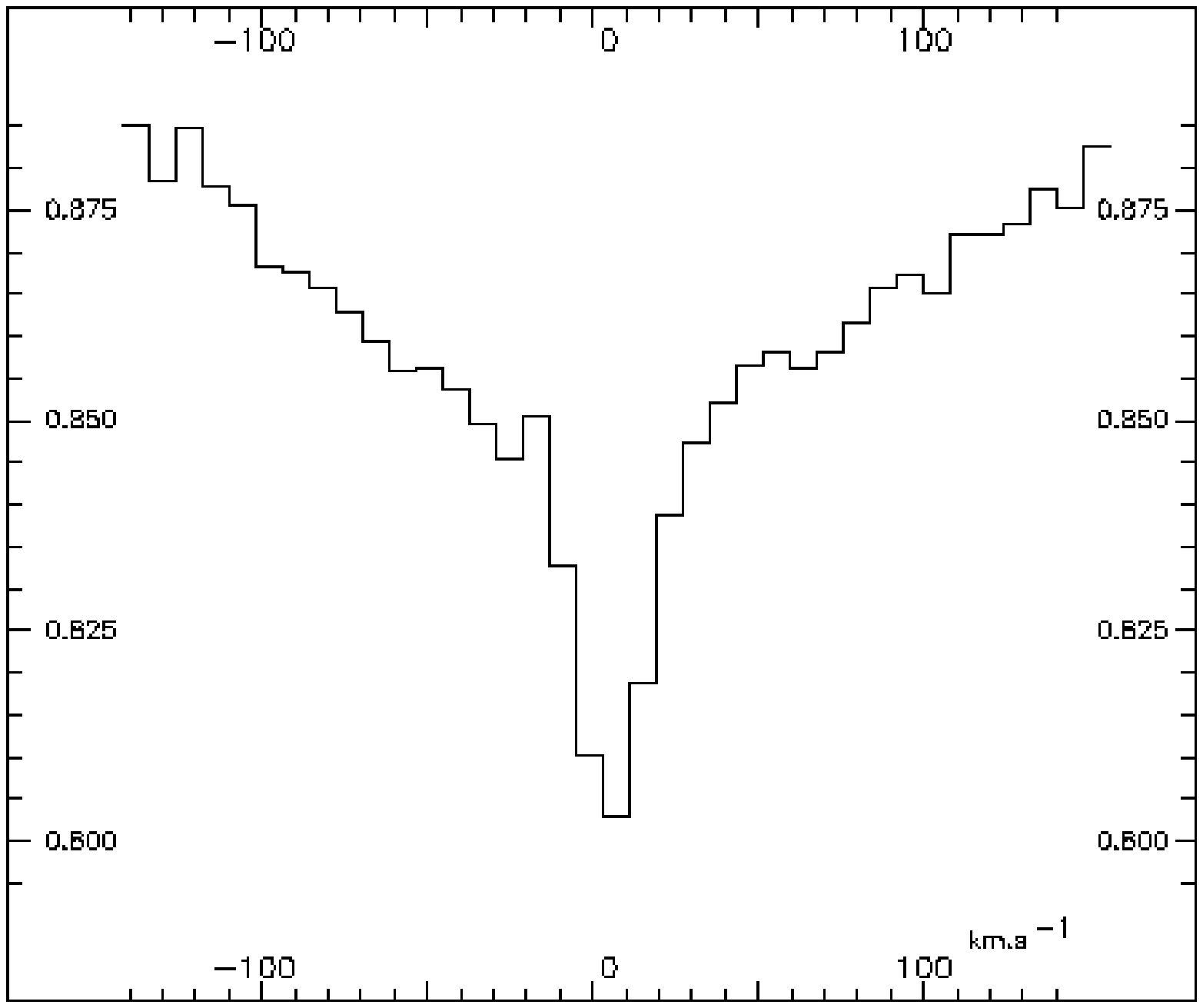}
\vspace{2mm}

{\bf Fig. 1e.} as Fig. 1a for HD 34787 2/11/95.
\end{figure}

\begin{figure}
\epsfxsize=8.8cm
\leavevmode\epsffile{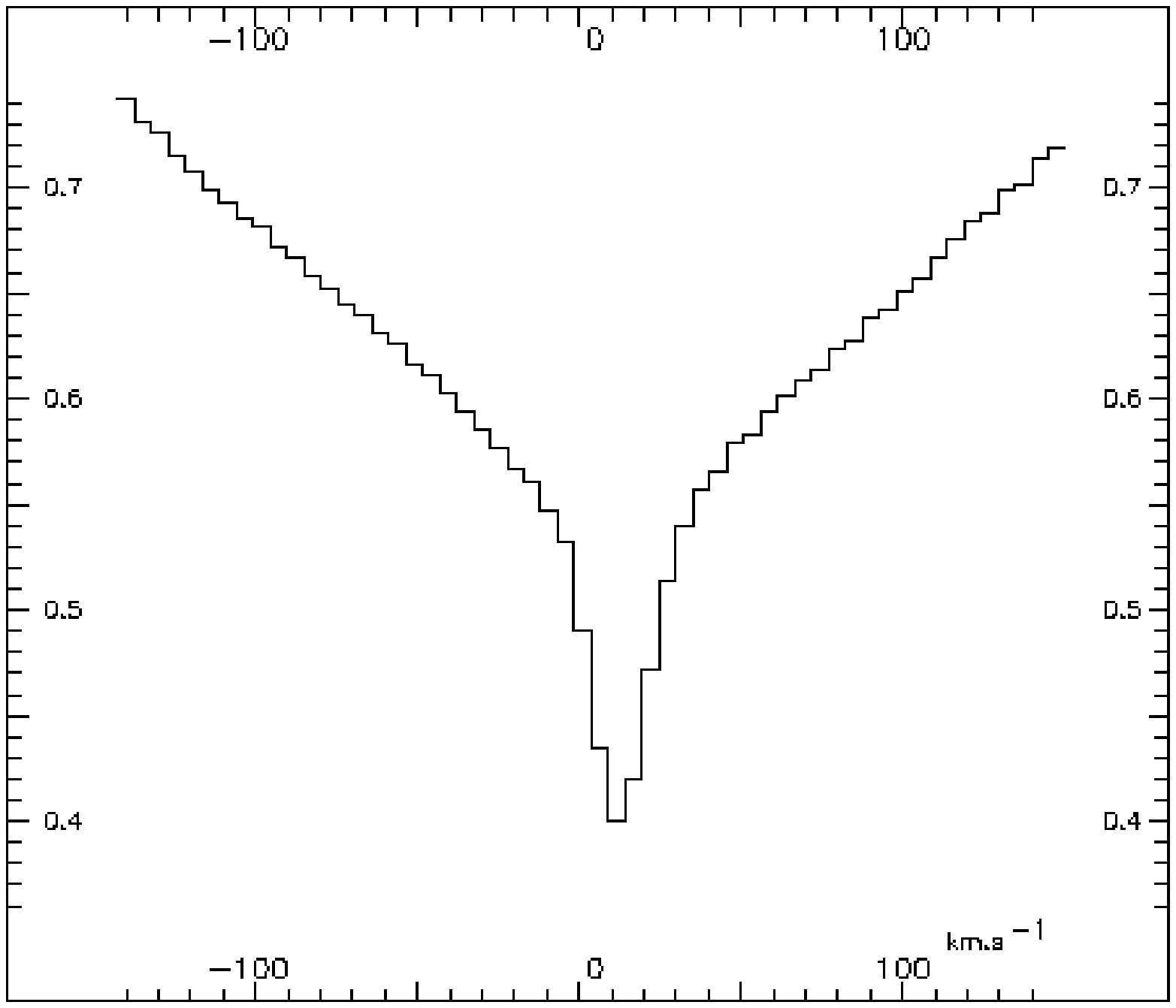}
\vspace{2mm}

{\bf Fig. 1f.} as Fig. 1a for HD 38545.
\end{figure}

\newpage
\begin{figure}
\epsfxsize=8.8cm
\leavevmode\epsffile{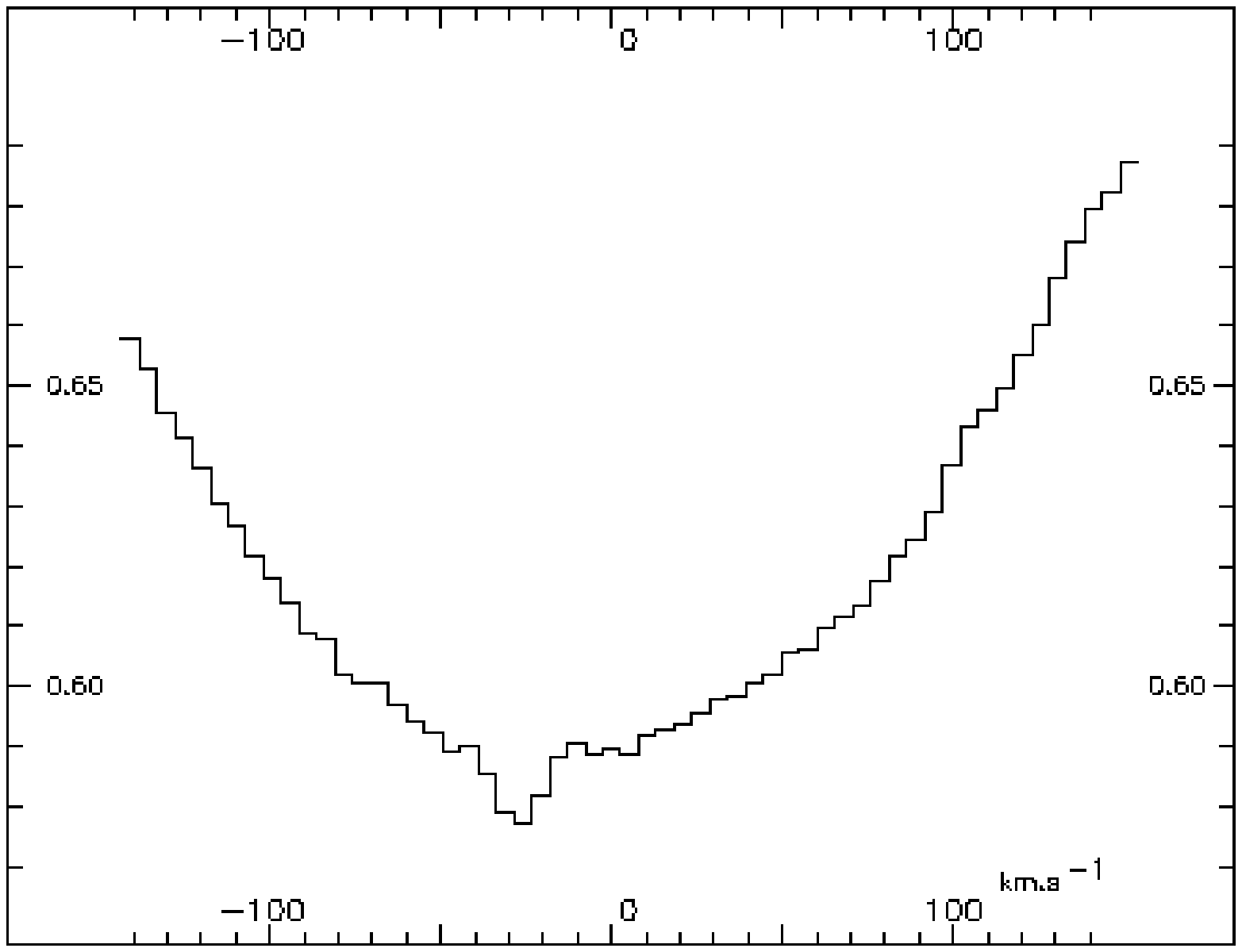}
\vspace{2mm}

{\bf Fig. 1g.} as Fig. 1a for HD 141851.
\end{figure}

\begin{figure}
\epsfxsize=8.8cm
\leavevmode\epsffile{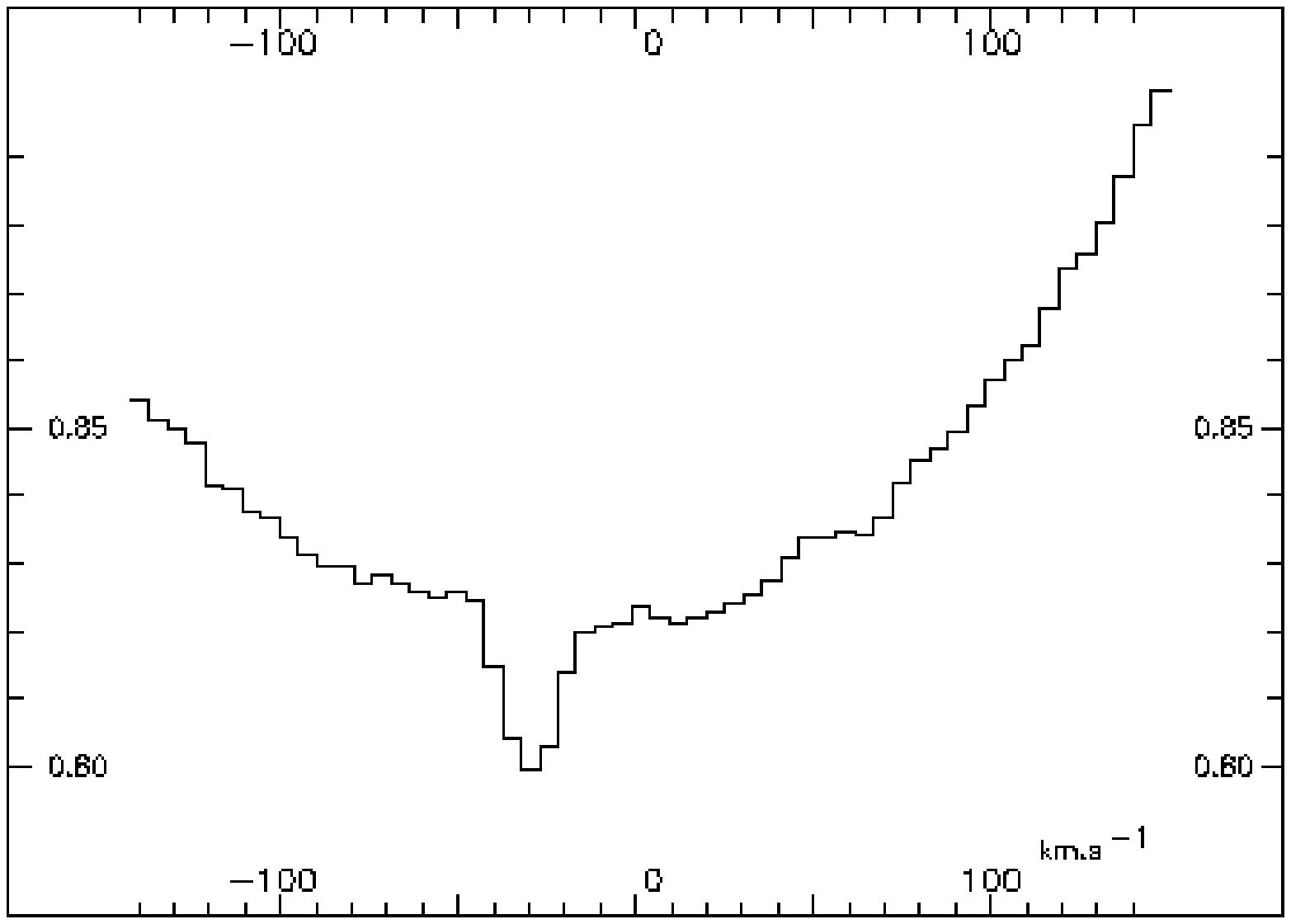}
\vspace{2mm}

{\bf Fig. 1h.} as Fig. 1a for HD 161868.
\end{figure}

\begin{figure}
\epsfxsize=8.8cm
\leavevmode\epsffile{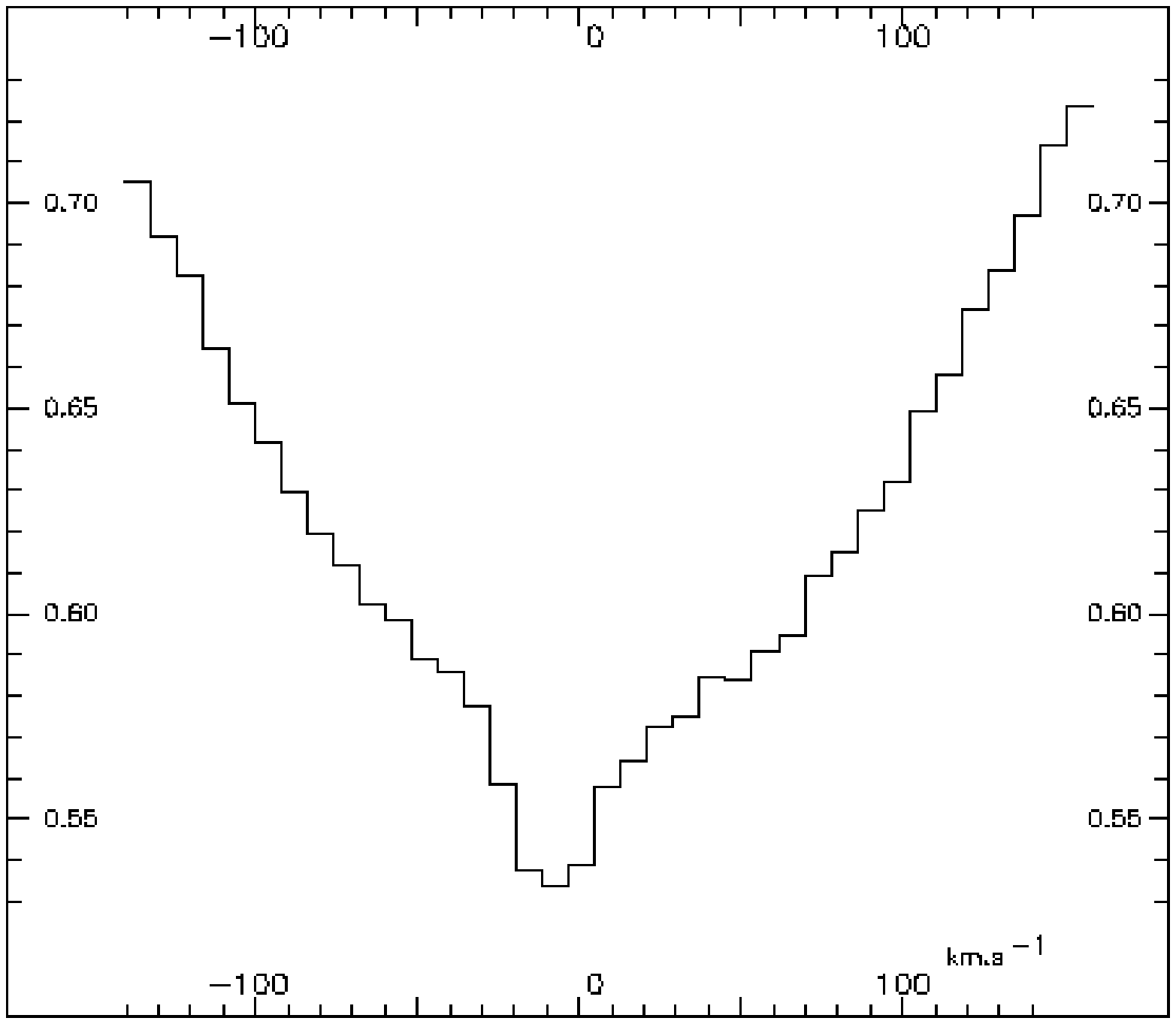} 
\vspace{2mm}

{\bf Fig. 1i.} as Fig. 1a for HD 217782.
\end{figure}

\begin{figure}
\epsfxsize=8.8cm
\leavevmode\epsffile{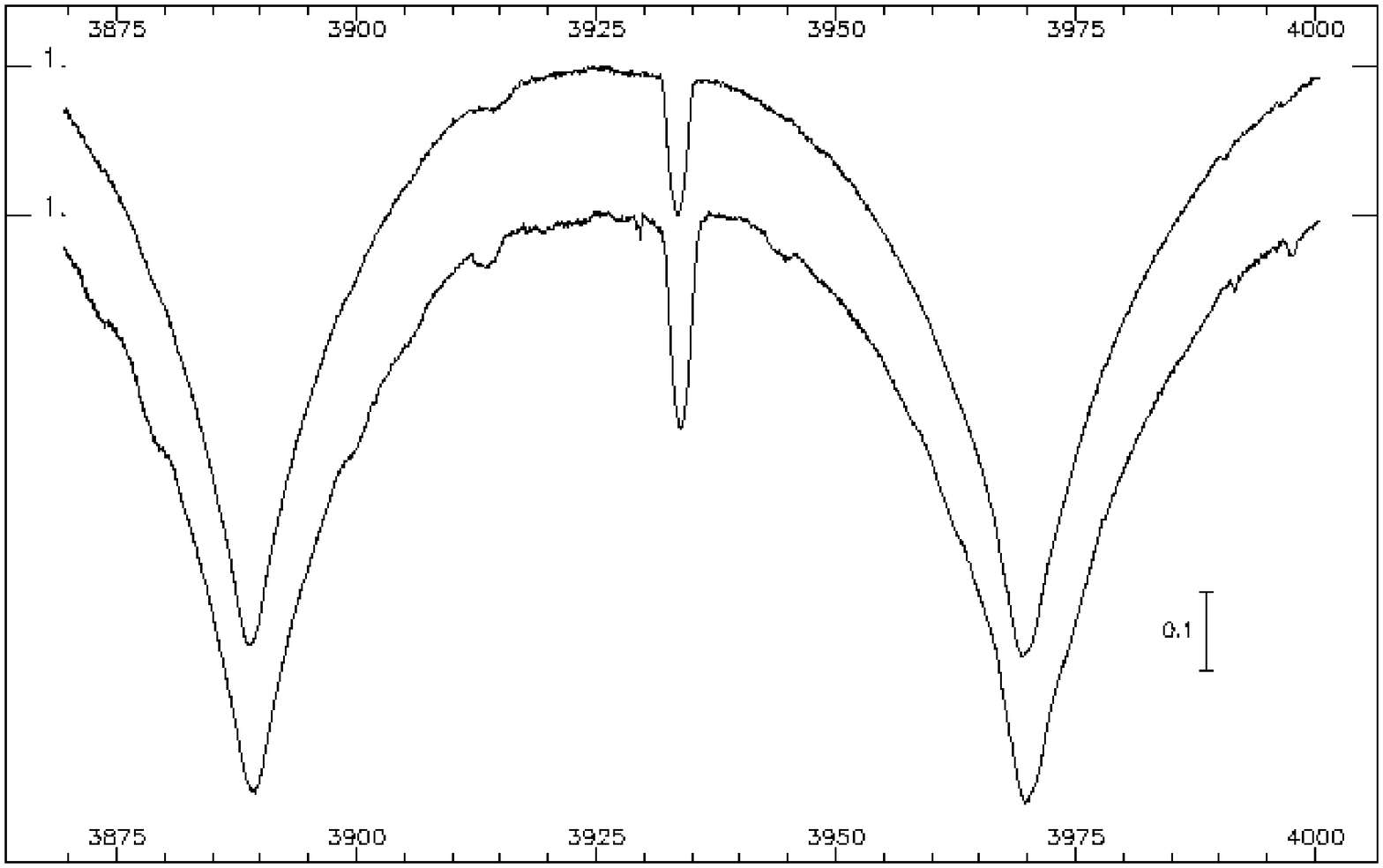}
\vspace{2mm}

{\bf Fig. 2.} Comparison of the spectrum of $\lambda$ Boo (upper spectrum) and $\pi^1$ Ori mear CaII-K.
\end{figure}
\newpage
\begin{figure}
\epsfxsize=8.8cm
\leavevmode\epsffile{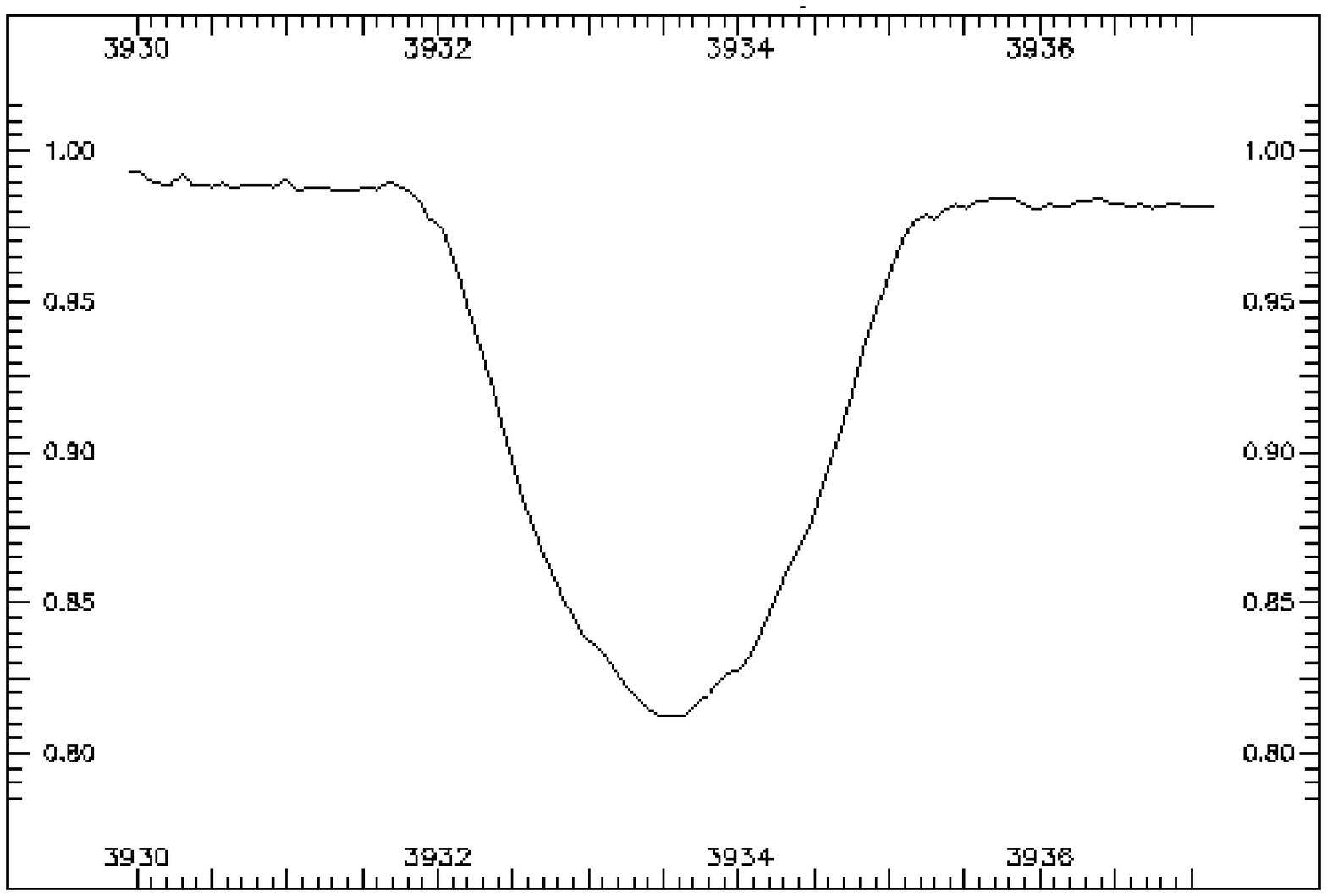}
\vspace{2mm}

{\bf Fig. 3.} CaII-K line profile of $\lambda$ Boo.
\end{figure}

\begin{figure}
\epsfxsize=8.8cm
\leavevmode\epsffile{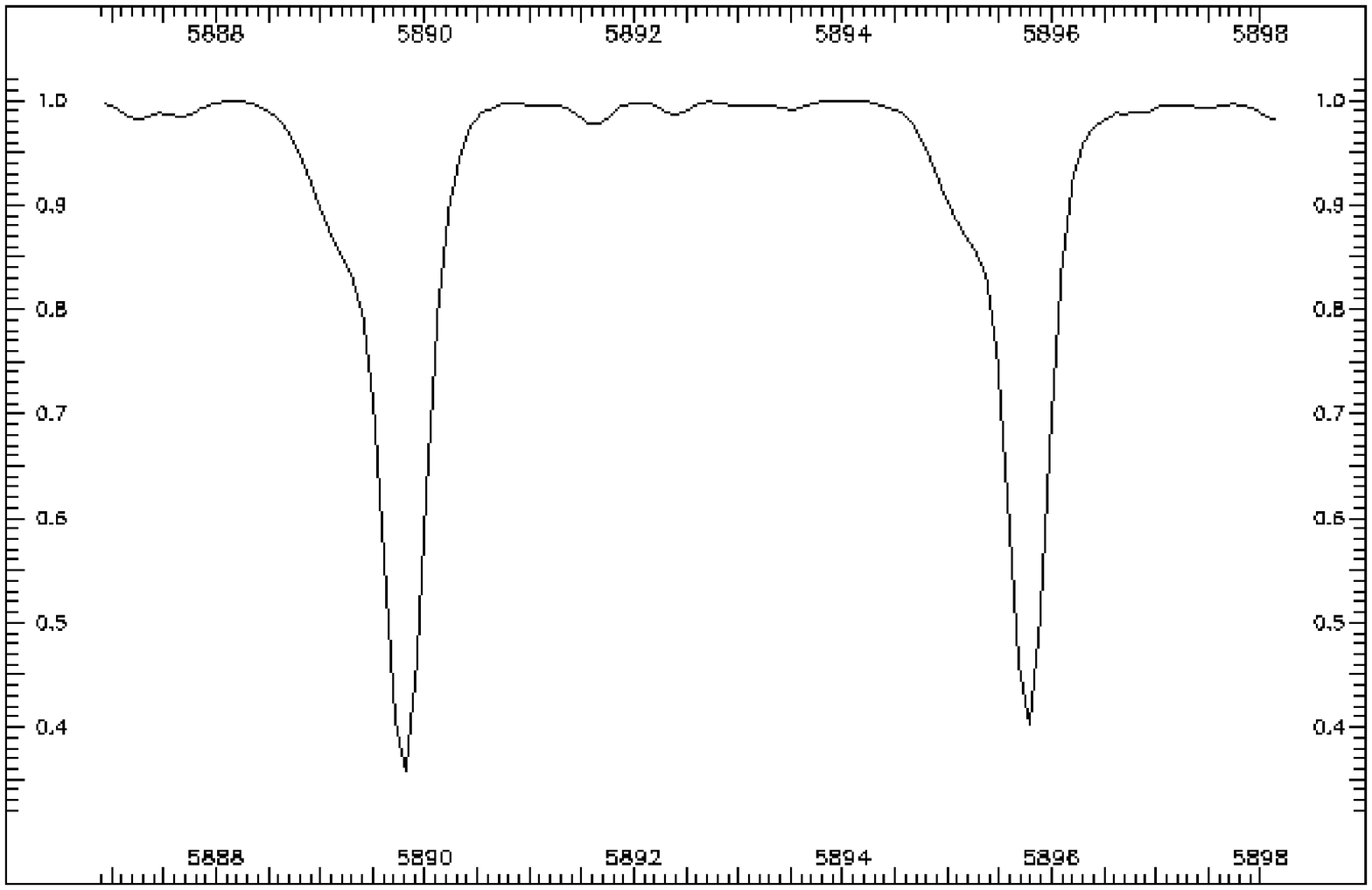}
\vspace{2mm}

{\bf Fig. 4.} Spectrum of HD 225180 near the NaI doublet.
\end{figure}

\end{document}